\documentclass[12pt]{article}
\usepackage{graphics}
\def\cA{{\cal A}}
\def\cJ{{\cal J}}
\def\cP{{\cal P}}

\usepackage{epsf}

\title{Photon Damping Caused by Electron-Positron Pair Production
in a Strong Magnetic Field}

\author{{M.V.~Chistyakov\thanks{E-mail address: mch@uniyar.ac.ru}, 
N.V.~Mikheev\thanks{E-mail address: mikheev@uniyar.ac.ru}}\\ [7mm]
{\small\it Division of Theoretical Physics, Department of Physics,}\\
{\small\it Yaroslavl State University, Sovietskaya 14,}\\ {\small\it
150000 Yaroslavl, Russian Federation}}

\begin{document}
\oddsidemargin 5mm
\maketitle

\begin{abstract}
Damping of an electromagnetic wave in a strong magnetic field is analyzed 
in the kinematic region near the
threshold of electron-positron pair production. Damping of the 
electromagnetic field is shown to be noticeably
nonexponential in this region. The resulting width of the 
photon $\gamma \to e^+ e^-$ decay is considerably smaller than
previously known results. 

\medskip

\end{abstract}

\newpage

The problem of propagation of electromagnetic
fields through an active medium is inherent in a variety
of physical phenomena. The birth and evolution of
supernova and neutron stars, where the matter density
can be on the order of nuclear density 
$\rho \simeq 10^{14} - 10^{15} g/cm^3$ and the temperature can 
achieve several tens of MeVs, are the largest scale and the most interesting
such phenomena. In addition to dense and hot matter, a
strong magnetic field, which can be several orders of
magnitude as high as so-called critical, or Schwinger,
value $B_e = m_e^2 / e \simeq 4.41 \cdot10^{13}G$
\footnote{We use the system of units where $\hbar = c = 1$.}, can be 
generated in the above-mentioned objects~\cite{bib:BG,bib:TD}. 
This strong magnetic field can induce new phenomena which can 
considerably affect the evolution of these astrophysical objects.
Electromagnetic-field damping caused by electron-positron pair production 
in an external magnetic field is
one of these phenomena. Recall that the $\gamma \to e^+ e^-$ 
process is kinematically forbidden in vacuum. The magnetic 
field changes the kinematics of charged particles,
electrons and positrons, allowing the production of an
electron-positron pair in the kinematic region 
$q_{\mbox {\tiny $\|$}}^2 = q_0^2 - q_3^2 \ge 4 m_e^2$,
where $q_0$ is the photon energy and $q_3$ is
the momentum component along the magnetic field
\footnote{Hereafter, we consider the magnetic field directed along the third
axis.}

In 1954, Klepikov~\cite{bib:K} examined the production of
an electron-positron pair by a photon in a magnetic
field and obtained the amplitude and width of the 
$\gamma \to e^+ e^-$  decay in the semiclassical approximation. Later,
the authors of~\cite{bib:S}--\cite{bib:Sh}  considered this process in the 
context of its astrophysical applications. It was pointed out
in~\cite{bib:B,bib:DH} that the use of the expression derived in~\cite{bib:K} 
for the width considerably overestimates the result in the
strong magnetic field limit. In this case, one should use
an exact expression for the width of one-photon production 
of a pair when electrons and positrons occupy
only the ground Landau level. However, it was found
that the expression for the decay width in the limit of
strong magnetic field has a root singularity at the point
$q_{\mbox {\tiny $\|$}}^2~=~4~m_e^2$. Shabad~\cite{bib:Sh} 
emphasized that this behavior
indicates that the decay width calculated in the 
perturbation theory cannot be treated as a damping 
coefficient. In this case, the damping coefficient is primarily
determined from the time evolution of the photon wave
function in the presence of a magnetic field. 
Shabad~\cite{bib:Sh} 
suggested that this dependence be obtained by solving
the dispersion equation with account taken of the 
vacuum polarization in a magnetic field with complex 
values of photon energy. In our opinion, this method has
several disadvantages. First, it is well known but rarely
mentioned that the dispersion equations with complex
energies have no solutions in the physical sheet. 
Solutions are in the nonphysical Riemannian sheets 
(analyticity region of the polarization operator), which are
generally infinite in number. As a result, the dispersion
equation has the infinite number of solutions with both
positive and negative imaginary parts of energy. The
physical status of these solutions requires a separate
investigation.

Shabad~\cite{bib:Sh} used the asymptotic form of the 
polarization operator near the pair production threshold and
erroneously treated it as a two-sheet complex function.
This circumstance led to the existence of two complex
conjugate solutions, one of which is physically 
meaningless because it has a positive imaginary part and,
therefore, provides exponentially increasing amplitude
of electromagnetic wave. Therefore, to obtain 
physically meaningful result, one should artificially discard
the redundant solutions.

Second, this approach cannot correctly describe the
substantially nonexponential damping in the near-threshold
region in a strong field.

Thus, damped electromagnetic waves in a magnetic
field cannot be completely described by solving the dispersion
equation.

In this work, we use a method that is applied in the
field theory at finite temperatures and in plasma physics
(see, e.g., ~\cite{bib:BdV}). It consists of the determination of a
retarded solution to the electromagnetic field equation
that includes an external source and takes into account
the vacuum polarization in a magnetic field. Time
damping of the electromagnetic wave is analyzed in a
uniform external magnetic field, whose intensity is the
largest parameter of the problem, $B_e \gg q^2, m_e^2$.
To describe the time evolution of electromagnetic
wave $\cA_{\alpha}(x)$ [$x_\mu = (t, {\bf x})$], we consider a linear response
of the system ($\cA_{\alpha}(x)$ and a vacuum polarized in magnetic
field) to an external source, which is adiabatically
turned on at $t = - \infty$ and turned off at $t = 0$. At $t > 0$, the
electromagnetic wave evolves independently. Thus, the
source is necessary for creating an initial state. For simplicity,
we consider the evolution of a monochromatic
wave. In this case, the source function should be taken
in the form
\begin{eqnarray}
\cJ_{\alpha}(x) = j_{\alpha}\,e^{i \,{\bf k} {\bf x}}\,
e^{ \varepsilon t}\, \theta(- t), \,\,\, \varepsilon \to 0^+.
\label{eq:1}
\end{eqnarray}
The time dependence of $\cA_{\alpha}(x)$ is determined by the
equation
\begin{eqnarray}
&& (g_{\alpha \beta} \, \partial_{\mu}^2  -
\partial_{\alpha}\partial_{\beta}) \, \cA_{\beta}(x) +
\int d^4 x'\, \cP_{\alpha \beta} (x - x') \, {\cal A}_{\beta}(x')
= {\cal J}_{\alpha}(x),
\label{eq:2}
\end{eqnarray}
where $\cP_{\alpha \beta} (x - x')$ is the photon polarization operator in
a magnetic field. We note that, for the source on the
right-hand side of Eq. (\ref{eq:2}) to be conserved, 
$\partial_\alpha \cJ_{\alpha} = 0$, 
the current
$j_\alpha$ must have the form 
$j_{\alpha} = (0, {\bf j}),\,\,{\bf j} \cdot {\bf k} = 0$. The evolution
of $\cA_{\alpha}(x)$ is described by the retarded solution
\begin{eqnarray}
\cA^R_{\alpha}(x) = \int d^4 x' \, G^R_{\alpha \beta} (x
- x')\, \cJ_{\beta}(x'),
\label{eq:3}
\end{eqnarray}
where $G^R_{\alpha \beta} (x - x')$ is the retarded Green's function,
which is defined through the commutator of the Heisenberg
operators of electromagnetic field as (see, e.g., ~\cite{bib:L&L})
\begin{eqnarray}
G^R_{\alpha \beta} (x - x') = - i \langle 0
|[\hat A_{\alpha} (x),\, \hat A_{\beta} (x')]|
0\rangle \,  \theta(t - t'),
\label{eq:4}
\end{eqnarray}
It is instructive to express this function in terms of the
causal Green's function
\begin{eqnarray}
G^C_{\alpha \beta} (x - x') = - i \langle 0
|{\mathbf T}\hat A_{\alpha} (x) \hat A_{\beta}
(x')| 0\rangle,
\label{eq:5}
\end{eqnarray}
by using the relationship
\begin{eqnarray}
G^R_{\alpha \beta}(x - x') = 2\, {\rm Re} \,
G^C_{\alpha \beta}(x - x')\, \theta(t -
t').
\label{eq:6}
\end{eqnarray}

In the presence of a magnetic field, it is convenient
to decompose Green's function (\ref{eq:5}) in the eigenvectors
of polarization operator~\cite{bib:Sh}:
\begin{eqnarray}
G^C_{\alpha \beta}(x) &=& \int \frac{d^4 q}{(2 \pi)^4}\,
G^C_{\alpha \beta}(q)\, e^{- i q x}
\label{eq:8} \\[2mm]%
G^C_{\alpha \beta}(q) &=& \sum_{\lambda = 1}^3
\frac{b_{\alpha}^{(\lambda)}
b_{\beta}^{(\lambda)}}{(b^{(\lambda)})^2} \, \frac{1} {q^2 -
\cP^{(\lambda)}(q) },
\label{eq:9}
\end{eqnarray}
where $\cP^{(\lambda)}(q)$ are the eigenvalues of polarization 
operator. The eigenvectors
 \begin{eqnarray}
 b_\alpha^{(1)} &=& (q \varphi)_\alpha, \nonumber \\
 b_\alpha^{(2)} &=& (q \tilde \varphi)_\alpha, \label{10} \\
 b_\alpha^{(3)} &=& q^2(q \varphi \varphi)_\alpha -
 (q\varphi \varphi q) q_\alpha,
\label{eq:10}
\end{eqnarray}
together with the 4-vector $q_\alpha$ form a complete 
orthogonal basis in the Minkowski 4-space. In Eqs. (\ref{eq:10}), 
$\varphi_{\alpha \beta} = F_{\alpha \beta} /B,\; {\tilde
\varphi}_{\alpha \beta} = \frac{1}{2} \varepsilon_{\alpha \beta
\mu \nu} \varphi_{\mu \nu}$ are dimensionless magnetic-field tensor 
and dual tensor, respectively, 
$ (q \varphi)_\alpha = q_\sigma
\varphi_{\sigma \alpha}$, \ $ (q \varphi \varphi q) = q_\alpha
\varphi_{\alpha \beta} \varphi_{\beta\sigma} q_\sigma$.
Substituting Eqs. (\ref{eq:1})  and
(\ref{eq:6}) into Eq. (\ref{eq:3}) and using Eqs. (\ref{eq:8}) 
and (\ref{eq:9}), we obtain
after simple integration the following result at $t > 0$:
\begin{eqnarray}
\cA^R_{\alpha} (x) =  \sum_{\lambda = 1}^3
V_\alpha^{(\lambda)}(x) = 2\, e^{ i\, {\bf k x}} \,
{\rm Re} \sum_{\lambda = 1}^3 \int \frac{d q_0}{2 \pi i} \,
\frac{b^{(\lambda)}_\alpha (b^{(\lambda)} j)/ (b^{(\lambda)})^2
\,\, e^{ - i \, q_0 t}}
{( q_0 - i
\varepsilon)\, (q_0^2 - {\bf k}^2 - \cP^{(\lambda)}(q) )},
\label{eq:11}
\end{eqnarray}
where $q_\alpha = (q_0, {\bf k})$. Note that the definition of the 
integral in Eq. (\ref{eq:11}) should be completed because the 
integrand can include singularities, which are due, on the
one hand, to zeros of its denominator and, on the other,
to the domain of its definition. To analyze these 
singularities, it is necessary to know the explicit form and
analytical properties of the eigenvalues $\cP^{(\lambda)}(q)$ of the
polarization operator, which was examined in detail in
a number of works. In the limit of strong magnetic field,
the functions $\cP^{(\lambda)}(q)$, which we are interested in, can be
borrowed, e.g., from ~\cite{bib:Sh,bib:TE,bib:MS} and represented as
[with the $O(1/B)$ accuracy]
\begin{eqnarray}
\cP^{(1)}(q) &\simeq& - \frac{\alpha}{3 \pi}\, 
q_{\mbox{\tiny $\bot$}}^2 -
q^2\, \Lambda(B, q^2), \label{P1}
\\[2mm]
\cP^{(3)}(q) &\simeq&  - \, q^2 \, \Lambda(B, q^2), 
\label{P3}
\\[2mm]
\cP^{(2)}(q) &\simeq& 
 - \frac{2 \alpha\, eB}{ \pi}  \, \Big (
\frac{1}{\sqrt{z (1 - z)}} 
\arctan  
\sqrt{\frac{z}{1 - z}} - 1 \Big ) -
%\nonumber\\
 q^2 \, \Lambda(B, q^2), \label{P2}
\end{eqnarray}
where 
$$
\Lambda(B, q^2) = \frac{\alpha}{3 \pi}\,\left[1.792 - \ln
(B/B_e)\right] + \pi(q^2),
$$
$$
z = q_{\mbox{\tiny $\|$}}^2 / 4 m_e^2, \,\,\, 
q_{\mbox {\tiny $\bot$}}^2 = q_1^2 + q_2^2,
$$
and $q^2 \, \pi(q^2)$ is the photon polarization operator in the
absence of a magnetic field. Note that the contribution
from the pole $q_{\mbox{\tiny $\|$}}^2 = 0$ that results from the 
normalization of the basis vectors $b^{(2)}_\alpha$ and 
$b^{(3)}_\alpha$ is nonphysical and,
taking into account explicit form (\ref{P1})--(\ref{P2}) of the 
polarization operator, can be removed by gauge transformation
after summation over polarizations. Thus, the contribution
to the solution can be made only by the poles
corresponding to the dispersion equation
\begin{equation}
q^2 - \cP^{(\lambda)}(q) = 0.
\label{eq:15}
\end{equation}
Using solution (\ref{eq:11}), one can demonstrate on the basis
of Eqs. (\ref{P1})--(\ref{P2}) that only two modes, $\lambda = 1$ and 
$\lambda =  2$ with the polarization vectors
\begin{equation}
\varepsilon _{\alpha}^{(1)} =
\frac{b_\alpha^{(1)}}{\sqrt{(b^{(1)})^2}} = \frac{ (q
\varphi)_{\alpha} } { \sqrt{ q^2_{\mbox{\tiny $\bot$}}}} ; \,\,
\varepsilon _{\alpha}^{(2)} =
\frac{b_\alpha^{(2)}}{\sqrt{(b^{(2)})^2}} = \frac{ (q \tilde
\varphi)_{\alpha} }{ \sqrt{ q^2_{\mbox{\tiny $\|$}}}}.
\label{eps}
\end{equation}
are physically meaningful for real photons
\footnote{
Modes with the polarizations $\varepsilon_{\alpha}^{(1)}$ and 
$\varepsilon_{\alpha}^{(2)}$
correspond to so-called
parallel ($\|$) and perpendicular ($\bot$) modes, respectively, in
the Adler notation~\cite{bib:Ad}.} 

A photon of the third, $\lambda = 3$, mode is nonphysical~\cite{bib:Sh}. 
Indeed, substitution of the expression for $\cP^{(3)}(q)$ into 
Eq. (\ref{eq:15})
gives the equation that has the only solution $q^2 = 0$.
Therefore, the contribution of the third mode to solution 
(\ref{eq:11}) is proportional to the total divergence and can
be eliminated by the corresponding gauge transformation.

In the limit of strong magnetic field, only the mode 
with the polarization vector $\varepsilon_{\mu}^{(2)}$ can decay into an 
electron-positron pair, because only the eigenvalue of the
polarization operator $\cP^{(2)}(q)$ (\ref{P2}) has the imaginary
part at $q_{\mbox{\tiny $\|$}}^2 \ge 4 m_e^2$. Therefore, to analyze 
time damping
of the electromagnetic field, it is sufficient to consider
only the term with $V^{(2)}_{\alpha} (x)$ in Eq. (\ref{eq:11}).

The further calculations can be simplified by going
over to the reference frame, where ${\bf k} = (k_1, k_2, 0)$, which
can always be done without disturbing the properties of
the external magnetic field. In this frame, $q_{\mbox{\tiny $\|$}}^2 = q_0^2$ 
and
the polarization vector of the second mode takes the
form $\varepsilon^{(2)}_\alpha = (0, 0, 0,- 1)$. As a result, 
$V^{(2)}_{\alpha} (x)$ is
expressed as
\begin{eqnarray}
V^{(2)}_{\alpha} (x) =  V^{(2)}_{\alpha} (0, {\bf x}) \,
\frac{{\rm Re} F(t)}{{\rm Re} F(0)},
\label{eq:17}
\end{eqnarray}
where
\begin{eqnarray}
F(t) = \int\limits_{C}\frac{d q_0}{2 \pi i} \, \frac{e^{ - i \, q_0 t}}{(
q_0 - i \varepsilon)\, (q_0^2 - {\bf k}^2 - {\cal P}^{(2)}(q) )},
\label{eq:18}
\end{eqnarray}
 
$$ V^{(2)}_{\alpha} (0, {\bf x}) = 2\, \varepsilon^{(2)}_\alpha
\,j_3\, e^{i\, {\bf k x}}\,{\rm Re} F(0).
$$

The path of integration $C$ in Eq. (\ref{eq:18}) is determined
by the analytical properties of $\cP^{(2)}(q)$ and is shown in
Fig. 1. The function $\cP^{(2)}(q)$ is analytical in the complex
plane $q_0$ with cuts along the real axis (see Fig. 1). In the
kinematic region $\vert q_0 \vert < 2 m_e$, the eigenvalue $\cP^{(2)}(q)$ is
real and Eq. (\ref{eq:15}) has real solutions which govern the
photon dispersion in this region.

For further analysis, it is convenient to transform the
path of integration to the path shown in Fig. 2. In this
case, the integral in Eq. (\ref{eq:18}) is represented as
\begin{eqnarray}
F(t) = F_{pole}(t) + F_{cut}(t),
\label{eq:19}
\end{eqnarray}
where the first term is determined by the residue at the
point $q_0 = \omega$, which is the solution to dispersion Eq.
(\ref{eq:15}). This term corresponds to the undamped solution
in the region $\omega < 2 m_e$~\cite{bib:Sh}. The second term determines
the time dependence of the electromagnetic field above
the threshold of electron-positron pair production and
has the form of the Fourier integral
\begin{eqnarray}
F_{cut}(t) &=& \int \limits_{- \infty}^{\infty} \frac{dq_0}{2 \pi}\,
F_{cut}(q_0)e^{- i q_0 t},
\label{eq:20}\\
F_{cut}(q_0) &=& 
\frac{2 \,\theta (q_0  -  2 m_e)\,I}
{q_0\,([ q_0^2 - {\bf k}^2 - R]^2 + I^2)},
\label{eq:21}
\end{eqnarray}
where
\begin{eqnarray}
R &\equiv& {\rm Re} \cP^{(2)}(q_0) =
\frac{ \alpha}{ \pi} \, eB \, \Big (
\frac{1}{\sqrt{z (z - 1)}} \ln  
\frac{\sqrt{z} + \sqrt{z-1}}{\sqrt{z} - \sqrt{z-1}} + 2 \Big ),
\label{eq:22}\\[2mm] 
I &\equiv&  - {\rm Im} \cP^{(2)}(q_0 + i \varepsilon) = 
\frac{\alpha \, eB}{\sqrt{z (z - 1)}}, \quad z = \frac{q_0^2}{4 m_e^2}.
%\nonumber\\
\label{eq:23}
\end{eqnarray}
Expressions (\ref{eq:20})--(\ref{eq:23}) together with 
Eq. (\ref{eq:17}) determine
the time evolution of the photon wave function above
the pair production threshold in a strong magnetic field.

Strictly speaking, because of the threshold behavior
of the Fourier transform $F_{cut}(q_0)$, time damping of the
function $F_{cut}(t)$ and, therefore, of the wave function
$\cA_\alpha(t)$ is nonexponential. However, in some characteristic
time interval (the inverse effective width of the
$\gamma \to e^+ e^-$ decay can naturally be chosen as such an
interval), the time dependence of the wave function can
approximately be represented as exponentially damping
harmonic oscillations:
\begin{equation}
\cA_\mu(t) \sim e^{- \gamma_{\mbox{\tiny eff}} \, t/2} \cos 
(\omega_{\mbox{\tiny eff}} t + \phi_0).
\label{eq:24}
\end{equation}
Here, $\omega_{\mbox{\tiny eff}}$ and $\gamma_{\mbox{\tiny eff}}$ 
are, respectively, the effective frequency
and width of photon decay, which should be
found by using Eqs. (\ref{eq:20})--(\ref{eq:23}) for each value of
momentum ${\bf k}$ to determine the effective photon dispersion
law above the threshold of electron-positron pair
production.
The quantity $\gamma_{\mbox{\tiny eff}}$, which governs the intensity of photon
absorption due to $e^+ e^-$ pair production in a magnetic
field, is important for astrophysical applications. The
absorption coefficient obtained from the $ \gamma \to e^+ e^-$ 
decay probability and containing a root singularity is
usually employed in astrophysics (see, e.g., ~\cite{bib:Bar}). 
Shabad~\cite{bib:Sh} pointed out that this leads to the overestimation
of the intensity of $e^+ e^-$-pair production. Our analysis
demonstrates that the calculation of the absorption
coefficient (decay width) by using the complex solution
in the second Riemannian sheet~\cite{bib:Sh}  also leads to a 
considerable overestimation in the near-threshold region,
as is seen from Figs. 3 and 4.

Nonexponential damping in the near-threshold
region is known for the processes in vacuum and 
matter~\cite{bib:Half, bib:Mats}. However, as far as we know, it has not been
considered in an external field so far. In contrast to 
vacuum or medium, the near-threshold effect in the 
magnetic field is kinematically enhanced due to the singular
behavior of the polarization operator in this field.
Therefore, this phenomenon is not only topical for
astrophysical application but is of conceptual interest.

We are grateful to S.S. Gershtein, G.P. Pron'ko,
V.V. Kiselev, and A.K. Likhoded for stimulating dis-
cussion and useful remarks.

This work was supported in part by the Ministry for
Education of the Russian Federation (project no. E00-
11.0-5) and the Russian Foundation for Basic Research
(project no. 01-02-17334).

\newpage

\thispagestyle{empty}

\begin{figure}[h]
\centerline{\includegraphics{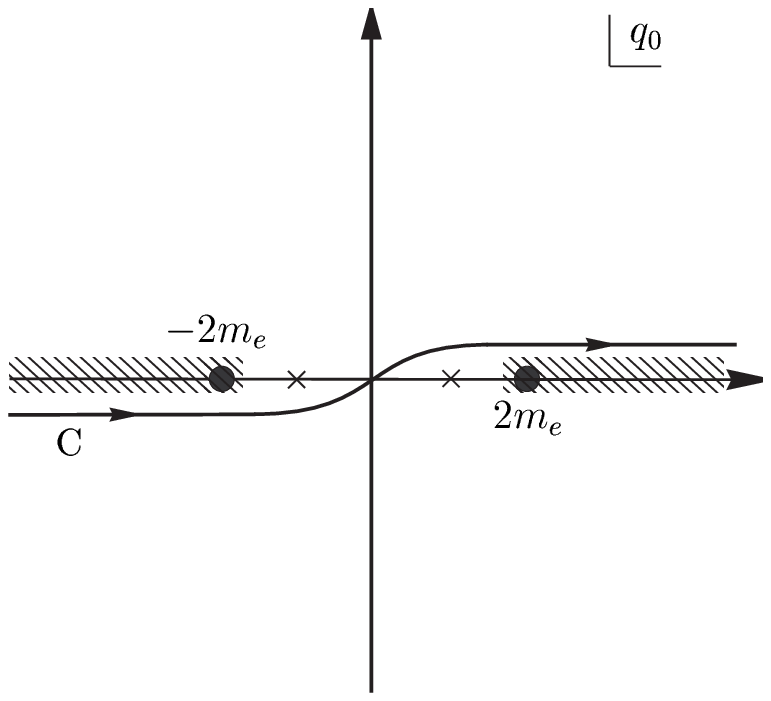}}
\caption{The path of integration $C$ in the complex $q_0$ plane.
The crosses are the poles corresponding to the real solutions
of dispersion Eq. (\ref{eq:15}). The shaded parts of the real axis are
cuts.}
\end{figure}

\newpage
\thispagestyle{empty}
\begin{figure}[h]
\centerline{\includegraphics{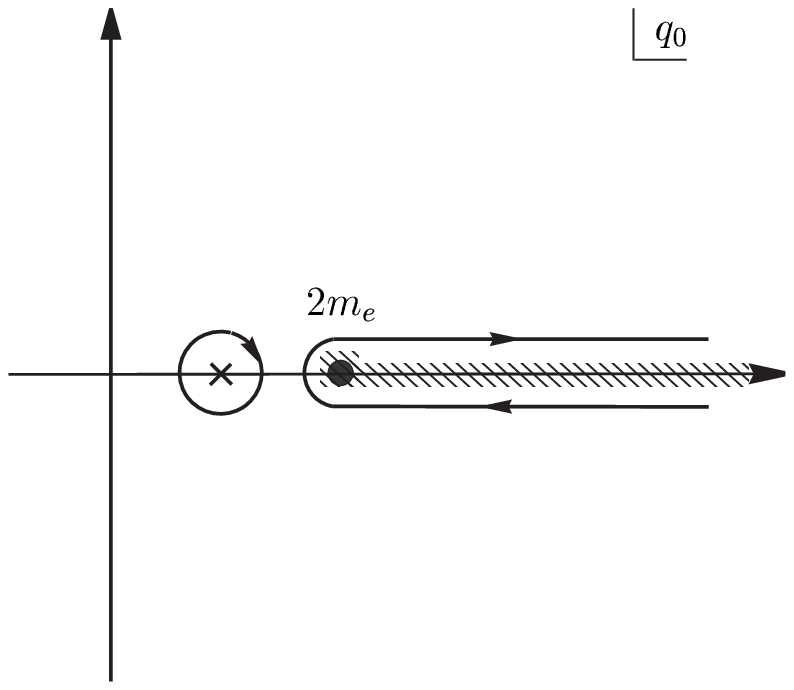}}
\caption{The path of integration $C$ after the transformation
allowing one to separate the pole $F_{pole}(t)$ and cut $F_{cut}(t)$
contributions.
}
\end{figure}

\newpage
\thispagestyle{empty}

\begin{figure}[h]
\centerline{\includegraphics{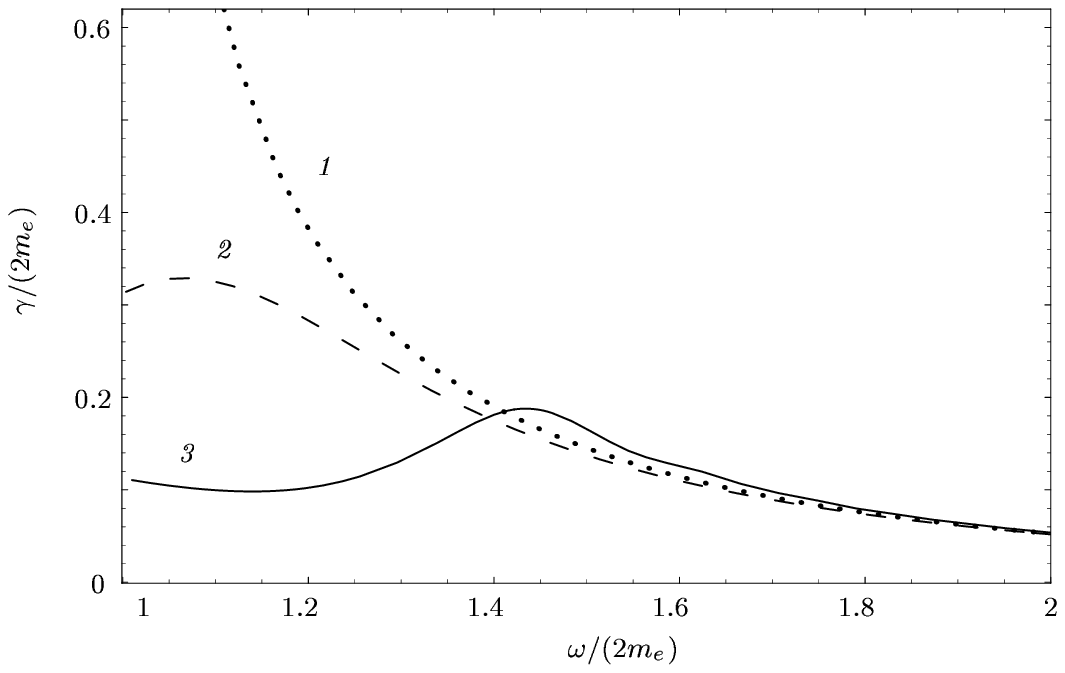}}
\caption{The frequency dependence of the $\gamma \to e^+ e^-$  decay
width in the near-threshold region for the magnetic field 
$B = 200 B_e$. Line ${\it 1}$ is the tree approximation including the root
singularity; line ${\it 2}$ is obtained from the complex solution of
the dispersion equation in the second Riemannian sheet~\cite{bib:Sh};
and line ${\it 2}$ is $\gamma_{\mbox{\scriptsize eff}}$ 
from approximation (\ref{eq:24}).}
\end{figure}

\newpage
\thispagestyle{empty}
\begin{figure}[h]
\centerline{\includegraphics{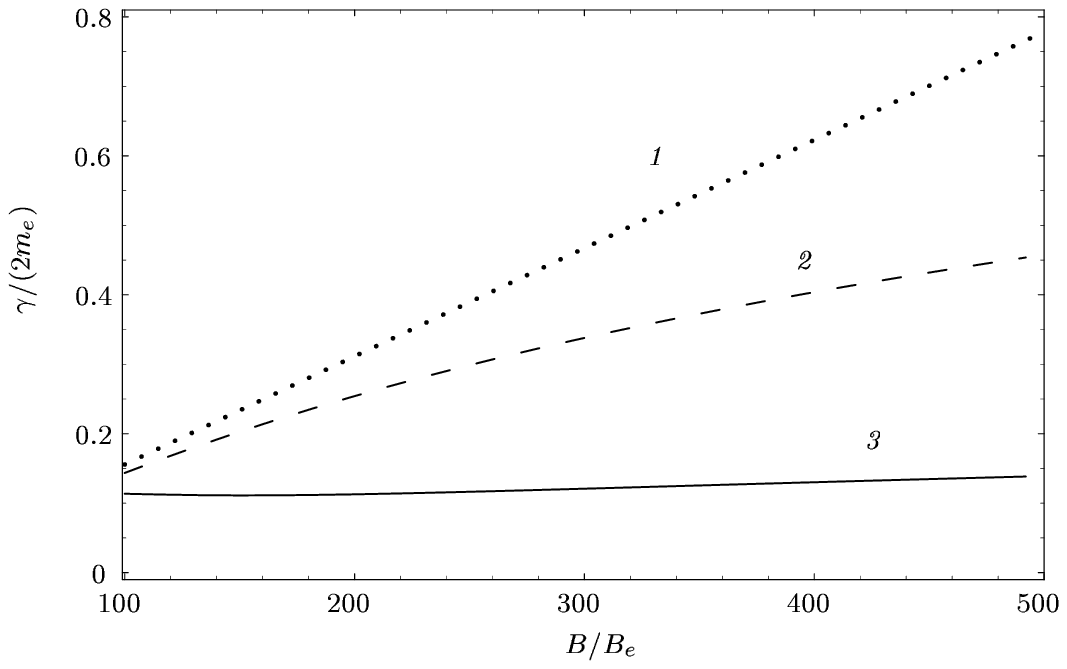}}
\caption{The decay width vs. the magnetic field for the 
frequency $\omega = 2.5 m_e$. The meaning of lines is the same as in
Fig. 3.}
\end{figure}

\end{document}